\PassOptionsToPackage{longnamesfirst}{natbib}

\documentclass[ecta,nameyear,draft]{econsocart}

\makeatletter
\let\numberlines@hook\relax
\let\set@numberlines@box\relax
\let\put@numberlines@box\relax
\makeatother

\RequirePackage[colorlinks,citecolor=blue,linkcolor=blue,urlcolor=blue,pagebackref]{hyperref}
\startlocaldefs

\theoremstyle{plain}

\newtheorem{proposition}{Proposition}

\newtheorem{lemma}{Lemma}

\theoremstyle{definition}
\newtheorem{definition}{Definition}

\newtheorem*{theorem*}{Blackwell's Theorem}

\usepackage[capitalise]{cleveref}
\crefname{figure}{Figure}{Figures}
\Crefname{figure}{Figure}{Figures}
\crefname{assumption}{Assumption}{Assumptions}
\crefname{footnote}{footnote}{footnotes}
\crefname{lemma}{Lemma}{Lemmas}

\usepackage{tikz}
\usepackage{tkz-euclide}
\usetikzlibrary{intersections}
\usepackage{pgfplots}
    \pgfplotsset{compat=1.15}
\usepackage{float}
\usepgfplotslibrary{fillbetween}
\definecolor{DodgerBlue}{RGB}{30,144,255}

\usepackage{standalone}
\usepackage{subcaption}
\usepackage{multirow}
\usepackage{nicematrix}
\usepackage{blkarray}
\usepackage{enumitem}

\usepackage{amssymb}

\usepackage[draft]{graphicx}

\newcommand{\td}{\widehat{\Delta}}
\newcommand{\bq}{\boldsymbol{q}}
\newcommand{\bp}{\boldsymbol{p}}

\newcommand{\bb}{\boldsymbol{b}}

\newcommand{\cp}{\mathcal{P}}

\newcommand{\ua}{\underline{a}}
\newcommand{\oa}{\overline{a}}

\newcommand{\hcp}{\widehat{\partial \cp}}
\newcommand{\mr}{\mathbb{R}}

\newcommand{\cb}{\mathcal{B}}

\newcommand{\cc}{\mathcal{C}}
\newcommand{\cd}{\mathcal{D}}
\newcommand{\hcc}{\widehat{\partial \cc}}

\newcommand{\cA}{\mathcal{A}}

\usepackage{caption} 
\endlocaldefs

\makeatletter
\def\copyright@text{}
\def\@oddhead{\hfil\small\thepage}
\def\@evenhead{\small\thepage\hfil}
\makeatother

\begin{document}

\begin{frontmatter}

\title{Quasi‑Concavity, Convexity of Optimal Actions, and the Local Single‑Crossing Property}
\runtitle{Ranking Statistical Experiments via the Linear Convex Order: Economic Applications}

\begin{aug}
%
%
%
\author[add1]{\fnms{Kailin}~\snm{Chen}}
\address[add1]{%
\orgdiv{January 19, 2026}\\
}
\end{aug}

\begin{funding}
Chen: Aalto University, Department of Economics, \href{mailto:kailin.chen@aalto.fi}{kailin.chen@aalto.fi}.

\end{funding}
\begin{abstract}

\end{abstract}


\end{frontmatter}

\section*{}

This note presents two results. First, it shows that under mild conditions, a decision problem is quasi-concave if the set of optimal actions is convex under every belief. Second, it shows that if a decision problem is quasi-concave, then it satisfies the local single crossing property after relabeling the states.

\section{Basic Setting}

A \textit{decision problem} $\{\cA,u\}$ consists of (\romannumeral 1) a set of actions $\cA$, and (\romannumeral 2) a payoff function $u:\cA\times\Theta\to\mathbb{R}$ (where $\Theta \subset \mathbb{R}$ denotes the state space). We let $a$ denote a generic action in $\cA$. The DM selects an action $a\in \cA$ and receives payoff $u(a,\theta)$ when the state is $\theta\in\Theta$. She is uncertain about the state and possesses a prior $\bq$. Upon observing a signal from an experiment, she updates her belief via Bayes’ rule and selects the action that maximizes her expected payoff.

In this note, we let $\Theta=\{\theta_0,\ldots,\theta_n\}$ and $\cA\subset \mathbb{R}$. For simplicity, we assume either (\romannumeral 1) $|\cA|<\infty$, or (\romannumeral 2) $\cA$ is a compact interval in $\mathbb{R}$ and $u(a,\theta)$ is differentiable as a function of $a$ for all $\theta$.

Given a belief $\bp$, we let $u(a,\bp):=\sum_{i=0}^{n}p_iu(a,\theta_i)$ denote the expected payoff of selecting action $a$. We let $\cA^*(\bp):=\operatorname*{arg\,max}_{a\in \cA}u(a,\bp)$ denote the set of optimal actions under $\bp$. 

\begin{definition}
\label{def:4}
    A decision problem $\{\cA,u\}$ is \textit{quasi-concave (QCC)} if $u(a,\bp)$ is quasi-concave as a function of $a$ for each $\bp\in\td_n$.
\end{definition}

\section{Equivalence between Quasi-concavity and Convexity}

If a decision problem is QCC, then for each $\bp$ the set of optimal actions $A^*(\bp)$ is convex relative to $A$. We now prove the converse.

\subsection{Finite Action Space}

Consider $A=\{a_0,\ldots,a_k\}$. Assume that
\begin{equation}
\label{eqn:38}
    \forall i\in\{0,\ldots,k\},\,\exists\,\bp(i)\in\td_n\ \text{such that}\ A^*(\bp(i))=\{a_i\}.
\end{equation}
By Theorem 1 in \citet{cheng2023dominance}, \eqref{eqn:38} is equivalent to requiring that no action is weakly dominated. Hence, \eqref{eqn:38} can be imposed without loss by iteratively eliminating weakly dominated actions. 

\begin{proposition}
\label{prop:12}
If \eqref{eqn:38} holds, then $\{A,u\}$ is QCC if $A^*(\bp)$ is convex relative to $A$ for each $\bp\in\td_n$.
\end{proposition}

\begin{proof}
For each $i\in\{0,\ldots,k-1\}$, let $\cp_i:=\{\bp\in\td_n \mid u(a_i,\bp)>u(a_j,\bp)\ \forall j>i\}$, $\hcp_i:=\partial\cp_i\cap\operatorname{int}\td_n$, and set $\cp_k:=\td_n$. For $i\neq j$, define $\cp(i,j):=\{\bp\in\td_n \mid u(a_i,\bp)=u(a_j,\bp)\}$. Since $u(a,\bp)$ is affine in $\bp$, each $\{\bp\mid u(a_i,\bp)>u(a_j,\bp)\}$ is an open halfspace, hence $\cp_0$ is convex and $\cp_0\cap\operatorname{int}\td_n$ is open. By \eqref{eqn:38} we have $\cp_0\neq\varnothing$; take $\bp(0)\in\cp_0\cap\operatorname{int}\td_n$.

For each $\bp\in\hcp_0$, continuity gives $u(a_0,\bp)\ge u(a_j,\bp)$ for all $j\ge1$, and since $\bp\notin\cp_0$ there exists $i\ge1$ with $u(a_0,\bp)=u(a_i,\bp)$. Thus $\{a_0,a_i\}\subseteq A^*(\bp)$; by convexity of $A^*(\bp)$ relative to $A$, $a_1\in A^*(\bp)$. Therefore $\hcp_0\subseteq\cp(0,1)\cap\operatorname{int}\td_n$.

Let $L_{i,j}(\bp):=u(a_i,\bp)-u(a_j,\bp)$ for $i\neq j$. Then $\cp(i,j)=\{\bp\in\operatorname{int}\td_n \mid L_{i,j}(\bp)=0\}$ is an affine hyperplane, and $\cp(0,1)$ partitions $\operatorname{int}\td_n$ into $\{L_{0,1}>0\}$ and $\{L_{0,1}<0\}$. Note that $\cp_0\subseteq\{L_{0,1}>0\}$. We claim $\{L_{0,1}>0\}\subseteq\cp_0$. Otherwise, pick $\bp^*\in\{L_{0,1}>0\}\setminus\cp_0$ and consider $\bp^t=(1-t)\bp(0)+t\bp^*$. Because $\cp_0$ is convex, $\cp_0\cap\operatorname{int}\td_n$ is open, and $\bp(0)\in\operatorname{int}\td_n$, there is $t\in(0,1)$ with $\bp^t\in\partial \cp_0$ and $\bp\in\operatorname{int}\td_n$. Hence. $\bp^t\in\hcp_0$. Then $\bp^t\in\cp(0,1)\cap\operatorname{int}\td_n$, so $L_{0,1}(\bp^t)=0$, but $L_{0,1}$ is affine and $L_{0,1}(\bp(0))>0$, $L_{0,1}(\bp^*)>0$, hence $L_{0,1}(\bp^t)>0$, a contradiction. Hence $\cp_0=\{L_{0,1}>0\}$ and $\hcp_0=\cp(0,1)\cap\operatorname{int}\td_n$.

We claim $A^*(\bp)=\{a_0,a_1\}$ for each $\bp\in\cp(0,1)\cap\operatorname{int}\td_n$. Equivalently, for each $i>1$ neither (i) $\cp(0,i)=\cp(0,1)$ nor (ii) $\cp(0,i)$ intersects $\cp(0,1)$ at some $\bp\in\operatorname{int}\td_n$ can occur. For (i): since $L_{0,j}$ is affine, $\cp(0,i)=\cp(0,1)$ implies $L_{0,i}=\alpha L_{0,1}$ with $\alpha\neq0$. Then $u(a_i,\bp)=(1-\alpha)u(a_0,\bp)+\alpha u(a_1,\bp)$ for all $\bp$, which makes either $a_i$ never uniquely optimal $(\alpha\in[0,1])$ or renders either $a_1$ $(\alpha>1)$ or $a_0$ $(\alpha<0)$ never uniquely optimal—contradicting \eqref{eqn:38}. Thus $\cp(0,i)\neq\cp(0,1)$. For (ii): let $\bp^\times\in\cp(0,1)\cap\cp(0,i)\cap\operatorname{int}\td_n$. Since the hyperplanes are distinct, there exists $v$ with $L_{0,i}(v)=0$ and $L_{0,1}(v)>0$. For small $t>0$, $\bp^t:=\bp^\times+tv\in\operatorname{int}\td_n$ satisfies $L_{0,i}(\bp^t)=0$ and $L_{0,1}(\bp^t)>0$, so $\bp^t\in\cp_0\cap\cp(0,i)$, contradicting $\cp_0=\{\bp\mid u(a_0,\bp)>u(a_j,\bp)\ \forall j\ge1\}$. Hence
\begin{equation}
\label{eqn:39}
u(a_1,\bp)>u(a_j,\bp)\quad\forall \bp\in\cp(0,1)\cap\operatorname{int}\td_n,\ \forall j>1.
\end{equation}

We claim $\{L_{0,1}>0\}\subseteq\{L_{1,j}>0\}$ for each $j>1$. Otherwise, take $j>1$ and $\bp'\in\{L_{0,1}>0\}$ with $L_{1,j}(\bp')\le0$. By \eqref{eqn:38} and a small perturbation if needed, pick $\bp(j)\in\operatorname{int}\td_n$ with $A^*(\bp(j))=\{a_j\}$. For each $\bp\in\{L_{0,1}>0\}$ we have $A^*(\bp)=\{a_0\}$ since $\{L_{0,1}>0\}=\cp_0$; furthermore, by continuity of $L_{0,1}$, for each $\bp\in\{L_{0,1}=0\}$ we have $a_0\in A^*(\bp)$. Hence $\bp(j)\in\{L_{0,1}<0\}$, so $L_{0,1}(\bp(j))<0$. Also $A^*(\bp(j))=\{a_j\}$ implies $L_{1,j}(\bp(j))<0$. Consider $\bp^t=(1-t)\bp(j)+t\bp'$. Since $\td_n$ is convex and $\bp(j)\in\operatorname{int}\td_n$, we have $\bp^t\in\operatorname{int}\td_n$ for all $t\in[0,1)$. Because $L_{0,1}(\bp(j))<0<L_{0,1}(\bp')$, there exists $t\in(0,1)$ with $L_{0,1}(\bp^t)=0$ and $\bp^t\in\cp(0,1)\cap\operatorname{int}\td_n$. By \eqref{eqn:39}, $L_{1,j}(\bp^t)>0$. Yet $L_{1,j}$ is affine and $L_{1,j}(\bp(j))<0$, $L_{1,j}(\bp')\le0$, hence $L_{1,j}(\bp^t)\le0$, a contradiction.

It follows that $\cp_1=\cap_{j>1}\{L_{1,j}>0\}$ and, since $\cp_0=\{L_{0,1}>0\}\subseteq\{L_{1,j}>0\}$ for all $j>1$, we have $\cp_0\subseteq\cp_1$. Repeating the same argument with $(0,1)$ replaced by $(1,2)$ (using \eqref{eqn:38} for $i=1$ and taking $\bp(1)\in\operatorname{int}\td_n$ by a small perturbation) yields $\hcp_1=\cp(1,2)\cap\operatorname{int}\td_n$ and $\cp_1=\{L_{1,2}>0\}$. Iterating for $i=0,\ldots,k-1$ gives the chain $\cp_0\subseteq\cp_1\subseteq\cdots\subseteq\cp_k$ (with $\cp_k=\td_n$) and identifies each region as $\cp_i=\{L_{i,i+1}>0\}$ for $i\in\{0,\ldots,k-1\}$.

Consequently, the adjacent‑comparison halfspaces are nested: $\{L_{0,1}>0\}\subseteq\{L_{1,2}>0\}\subseteq\cdots\subseteq\{L_{k-1,k}>0\}$. For each $i\in\{0,\ldots,k-1\}$ and every $\bp$, let $D_i(\bp):=L_{i,i+1}(\bp)=u(a_i,\bp)-u(a_{i+1},\bp)$. If $D_i(\bp)>0$, then $\bp\in\{L_{i,i+1}>0\}$ and, by nesting, for every $j>i$, we have $\bp\in\{L_{j,j+1}>0\}$ and $D_j(\bp)>0$. Conversely, if $D_i(\bp)<0$, then $\bp\notin\{L_{i,i+1}>0\}$. For every $j<i$, since $\{L_{j,j+1}>0\}\subseteq\{L_{i,i+1}>0\}$, it follows that $\bp\notin\{L_{j,j+1}>0\}$ and $D_j(\bp)\le0$. Hence, the sequence $i\mapsto u(a_i,\bp)$ is unimodal for each $\bp$, and $\{A,u\}$ is QCC.
\end{proof}

\subsection{Continuous Action Space}

Consider $A=[\ua,\oa]$ and suppose $u(a,\theta)$ is differentiable in $a$ for all $\theta$. Assume that
\begin{equation}
\label{eqn:40}
\exists\,\bp(\ua),\bp(\oa)\in\td_n \ \text{such that}\ \ua\in A^*(\bp(\ua)),\ \oa\in A^*(\bp(\oa)).
\end{equation}
Condition \eqref{eqn:40} can be imposed \emph{without loss} by iteratively eliminating strictly dominated actions from both ends of $A$. Furthermore, since $u(a,\bp)$ is continuous in $(a,\bp)$ and $A$ is compact, Berge’s maximum theorem implies that $A^*(\bp)$ is nonempty, compact‑valued, and upper hemicontinuous in $\bp$. Let $S:=\{(\bp,a)\in\td_n\times A \mid a\in A^*(\bp)\}$ be the graph of $A^*$. Then $S$ is compact, so its projection $K:=\{a\in A \mid \exists\,\bp\in\td_n \ \text{with}\ a\in A^*(\bp)\}$ is nonempty and compact. Hence the endpoints $\ua^*:=\min K$ and $\oa^*:=\max K$ are attained at some $\bp(\ua^*),\bp(\oa^*)\in\td_n$, and replacing $A$ by $[\ua^*,\oa^*]$ yields \eqref{eqn:40}.

\begin{proposition}
\label{prop:13}
If \eqref{eqn:40} holds, then $\{A,u\}$ is QCC provided $A^*(\bp)$ is convex for each $\bp\in\td_n$.
\end{proposition}

We now prove \cref{prop:13}. For simplicity, we assume $A^*(\bp)$ is a singleton for each $\bp\in\td_n$ and write $a^*(\bp)$ for the unique optimizer. By Berge’s maximum theorem, $a^*(\bp)$ is continuous in $\bp$.

For $a\in(\ua,\oa)$ define the sublevel region
\[
\cc(a):=\{\bp\in\td_n \mid a^*(\bp)<a\},
\]
its interior boundary
\[
\hcc(a):=\partial\cc(a)\cap\operatorname{int}\td_n,
\]
and the first–order (FOC) hyperplane at $a$
\[
\cd(a):=\Bigl\{\bp\in\td_n \ \Bigm|\ \frac{\partial u(a,\bp)}{\partial a}=0\Bigr\}.
\]
Note that $\cd(a)$ is an affine hyperplane intersected with $\td_n$. Because $a^*(\cdot)$ is continuous, $\cc(a)$ is (relatively) open in $\td_n$, equivalently $\cc(a)\cap\operatorname{int}\td_n$ is open.

Fix $a\in(\ua,\oa)$ and $\bp\in\hcc(a)$, continuity of $a^*(\cdot)$ forces $a^*(\bp)=a$. As $a$ is interior in $A$, optimality at $\bp$ yields the first–order condition $\partial u(a,\bp)/\partial a=0$. Hence
\begin{equation}
\label{eqn:41}
\hcc(a)\subseteq \cd(a)\cap\operatorname{int}\td_n.
\end{equation}

Recalling the discrete case in \cref{prop:12} (where we first established $\hcp_0\subseteq\cp(0,1)\cap\operatorname{int}\td_n$ and then upgraded to equality), our next goal is the continuous analogue:
\begin{equation}
\label{eqn:42}
\hcc(a)=\cd(a)\cap\operatorname{int}\td_n.
\end{equation}
Note that \eqref{eqn:42} is intuitive: if the interior boundary of $\cc(a)$ is contained in a hyperplane, then on the interior it must coincide with that hyperplane; otherwise the interior boundary would have “holes.” We formalize this via the following lemma.

\newcommand{\co}{\mathcal{O}}
\newcommand{\cs}{\mathcal{S}}
\newcommand{\oco}{\overline{\co}}
\newcommand{\hcd}{\widehat{\cd}}

\begin{lemma}
\label{lem:5}
Let $\co\subset\mr^k$ be nonempty, open, convex, and connected. Let $\cc\subseteq\oco$ be relatively open in $\oco$, and set $\cc':=\cc\cap\co$ and $\hcc:=\partial\cc\cap\co$. Let $\cd\subset\mr^k$ be an affine hyperplane and put $\hcd:=\cd\cap\co$. If \textup{(i)} $\hcc\subseteq\hcd$, \textup{(ii)} $\cc'\neq\varnothing$, and \textup{(iii)} $\co\setminus\cc'$ has nonempty interior in $\co$, then $\hcc=\hcd$.
\end{lemma}

\begin{proof}
Set $\cb:=\co\setminus\hcd$. Since $\co$ is open, $\cc'$ is open in $\co$, and
\[
\hcc=\partial\cc\cap\co=\partial_{\co}(\cc').
\]
Because $\cb$ is open in $\co$, for any $\cs\subseteq\co$ we have
\[
\partial_{\cb}(\cs\cap\cb)=\partial_{\co}\cs\cap\cb.
\]
Applying this with $\cs=\cc'$ yields
\[
\partial_{\cb}(\cc'\cap\cb)=\partial_{\co}\cc'\cap\cb=\hcc\cap\cb.
\]
Since $\hcc\subseteq\hcd$, we have $\hcc\cap\cb=\varnothing$, hence
$\partial_{\cb}(\cc'\cap\cb)=\varnothing$. Therefore $\cc'\cap\cb$ is both open and
closed (clopen) in $\cb$.

\emph{Case 1: $\hcd=\varnothing$.} Then $\cb=\co$ is connected, so $\cc'\cap\cb\in\{\varnothing,\co\}$ and thus $\cc'\in\{\varnothing,\co\}$, contradicting \textup{(ii)}–\textup{(iii)} (the complement would have empty interior).

\emph{Case 2: $\hcd\neq\varnothing$.} Then $\cb=\co\setminus\hcd$ has exactly two connected components, say $\cb_1$ and $\cb_2$ (the two open halfspaces determined by $\cd$, intersected with $\co$). Since $\cc'\cap\cb$ is clopen in $\cb$, it must be one of $\varnothing$, $\cb_1$, $\cb_2$, or $\cb_1\cup\cb_2(=\cb)$.

- If $\cc'\cap\cb=\varnothing$, then $\cc'\subseteq\hcd$. Because $\cc'$ is open in $\co$ and $\hcd$ has empty interior in $\co$, it follows that $\cc'=\varnothing$, contradicting \textup{(ii)}.

- If $\cc'\cap\cb=\cb$, then $\co\setminus\cc'\subseteq\hcd$, whose interior in $\co$ is empty—contradicting \textup{(iii)}.

- Thus $\cc'\cap\cb=\cb_i$ for some $i\in\{1,2\}$. Write $K:=\cc'\cap\hcd$, so $\cc'=\cb_i\cup K$. Since $\cc'$ is open in $\co$, $K$ must be empty (any point of $K\subset\hcd$ has neighborhoods in $\co$ meeting both components). Hence $K=\varnothing$ and $\cc'=\cb_i$. Finally,
\[
\hcc=\partial_{\co}\cc'=\partial_{\co}\cb_i=\hcd,
\]
as claimed.
\end{proof}

We apply \cref{lem:5} with $\co=\operatorname{int}\td_n$, $\cc=\cc(a)$, and $\cd=\cd(a)$. 
Furthermore, by \eqref{eqn:40} and the continuity of $a^*(\cdot)$, both $\cc(a)$ and 
$\cc^c(a):=\{\bp\in\td_n \mid a^*(\bp)>a\}$ are open in $\td_n$ and nonempty. Hence 
$\cc(a)\cap\operatorname{int}\td_n$ and $\cc^c(a)\cap\operatorname{int}\td_n$ are nonempty and open. 
Therefore $\operatorname{int}\td_n\setminus\bigl(\cc(a)\cap\operatorname{int}\td_n\bigr)$ has nonempty interior in 
$\operatorname{int}\td_n$. Thus, by \cref{lem:5}, we establish
\begin{equation}
\label{eqn:42}
\hcc(a)=\cd(a)\cap\operatorname{int}\td_n.
\end{equation}

By \eqref{eqn:42}, we obtain
\begin{equation}
\label{eqn:43}
\frac{\partial u(a,\bp)}{\partial a}=0 \ \iff\ a\in A^*(\bp)\quad \forall\,\bp\in\td_n,\ a\in(\ua,\oa).
\end{equation}
The implication “$\Leftarrow$” is immediate. For “$\Rightarrow$”: if $\bp\in\cd(a)\cap\operatorname{int}\td_n$, then 
$\bp\in\hcc(a)$ by \eqref{eqn:42}, hence $a^*(\bp)=a$. For $\bp\in\partial\td_n$, pick 
$\bp^m\in\cd(a)\cap\operatorname{int}\td_n$ with $\bp^m\to\bp$; continuity of $a^*(\cdot)$ yields $a^*(\bp)=a$.

Now fix $\bp$ with $a^*(\bp)\in(\ua,\oa)$. We claim that for every $a\in(\ua,a^*(\bp))$ we have 
$\partial u(a,\bp)/\partial a>0$. Otherwise, take $a'<a^*(\bp)$ with $\partial u(a',\bp)/\partial a\le 0$. 
If $\partial u(a',\bp)/\partial a=0$, then by \eqref{eqn:43} we must have $a'=a^*(\bp)$, a contradiction. 
If $\partial u(a',\bp)/\partial a<0$, then since $a^*(\bp)$ maximizes $u(\cdot,\bp)$ we have 
$u(a^*(\bp),\bp)>u(a',\bp)$. By the Mean Value Theorem there exists $a''\in(a',a^*(\bp))$ with
\[
\frac{\partial u(a'',\bp)}{\partial a}=\frac{u(a^*(\bp),\bp)-u(a',\bp)}{a^*(\bp)-a'}>0.
\]
By Darboux’s theorem (derivatives have the intermediate value property), there exists 
$a^0\in(a',a'')$ with $\partial u(a^0,\bp)/\partial a=0$, and then \eqref{eqn:43} gives 
$a^0=a^*(\bp)$, again a contradiction. Therefore $\partial u(a,\bp)/\partial a>0$ for all 
$a\in(\ua,a^*(\bp))$. A symmetric argument shows that for every $a\in(a^*(\bp),\oa)$ we have 
$\partial u(a,\bp)/\partial a<0$. Hence $u(a,\bp)$ is unimodal on $(\ua,\oa)$ and, by continuity, on $[\ua,\oa]$, 
establishing quasi-concavity.

Futhermore, consider the endpoints $\ua,\oa$ (derivatives interpreted from within $A$). 
If $a^*(\bp)=\ua$, then $\partial u(\ua,\bp)/\partial a\le 0$. We now show that for every $a'\in(\ua,\oa)$ 
we cannot have $\partial u(a',\bp)/\partial a\ge 0$. Indeed, if $\partial u(a',\bp)/\partial a=0$, then by 
\eqref{eqn:43} we would have $a^*(\bp)=a'$, a contradiction. If $\partial u(a',\bp)/\partial a>0$, then, since 
$u(a',\bp)\le u(\ua,\bp)$ by optimality of $\ua$, the Mean Value Theorem yields $a''\in(\ua,a')$ with
\[
\frac{\partial u(a'',\bp)}{\partial a}=\frac{u(a',\bp)-u(\ua,\bp)}{a'-\ua}\le 0.
\]
If equality holds, $\partial u(a'',\bp)/\partial a=0$ and \eqref{eqn:43} gives $a^*(\bp)=a''>\ua$, a contradiction; hence 
$\partial u(a'',\bp)/\partial a<0$. By Darboux, there exists $a^)\in(a'',a')$ with 
$\partial u(a^0,\bp)/\partial a=0$, again contradicting \eqref{eqn:43}. Therefore 
$\partial u(a,\bp)/\partial a<0$ for all $a\in(\ua,\oa)$ when $a^*(\bp)=\ua$. The case $a^*(\bp)=\oa$ is analogous 
(with the inequalities reversed).

Therefore, $u(a,\bp)$ is unimodal for every $\bp$, establishing QCC.

Finally, when $A^*(\bp)$ is not always a singleton, for each $\bp$ there exist $\alpha(\bp)\le\beta(\bp)$ such that $A^*(\bp)=[\alpha(\bp),\beta(\bp)]$. By Berge’s maximum theorem, $A^*(\cdot)$ is nonempty, compact-valued, and upper hemicontinuous; hence (i) $\alpha(\cdot)$ is upper semicontinuous and (ii) $\beta(\cdot)$ is lower semicontinuous. Repeating the argument above—with $\cc(a):=\{\bp\in\td_n \mid \alpha(\bp)<a\}$ for the “left” side and $\cc^\uparrow(a):=\{\bp\in\td_n \mid \beta(\bp)>a\}$ for the “right” side—yields that $u(a,\bp)$ is weakly increasing in $a$ on $[\ua,\alpha(\bp)]$, constant on $[\alpha(\bp),\beta(\bp)]$, and weakly decreasing on $[\beta(\bp),\oa]$. Therefore $u(a,\bp)$ is unimodal in $a$ for every $\bp$, establishing QCC.

\section{Relation to Local Single Crossing Property}

A vector $\bb=(b_0,\ldots,b_n)\in\mathbb{R}^{n+1}$ is \emph{quasi-monotone} if there exists $k\in\{1,\ldots,n\}$ such that $b_i\le 0$ for all $i<k$ and $b_i\ge 0$ for all $i\ge k$.

Consider the decision problem $\{\cA,u\}$ with $\cA\subseteq\mathbb{R}$. Recall that we assume either (i) $\cA=\{a_0,\ldots,a_k\}$, or (ii) $\cA=[\underline a,\overline a]$ and $u(a,\theta)$ is differentiable in $a$ for all $\theta$. In case (i), define $\Delta u(a_i,\theta):=u(a_i,\theta)-u(a_{i-1},\theta)$ for each $i\ge 1$ and let $\Delta u(a_0,\theta)\equiv 0$. In case (ii), define $\Delta u(a,\theta):=\partial u(a,\theta)/\partial a$ for each $a\in \cA$. The decision problem is \emph{locally-single-crossing (LSC)} if, for each $a\in \cA$, the vector $\big(\Delta u(a,\theta_0),\ldots,\Delta u(a,\theta_n)\big)$ is quasi-monotone. Note that LSC is implied by the single-crossing property of \citet{milgrom1994monotone} and the interval dominance order property of \citet{quah2009comparative}, while the converse need not hold.

\begin{proposition}
\label{prop:7}
    If a decision problem is QCC, then there exists a bijection $\beta:\Theta\to\Theta$ such that the problem is LSC on the state space relabeled by $\beta$.
\end{proposition}

\begin{proof}
    For each $\theta\in\Theta$, define $a^*(\theta):=\min\{\arg\max_{a\in A} u(a,\theta)\}$. Relabel the states by nondecreasing $a^*(\theta)$, breaking ties arbitrarily; let $\beta:\Theta\to\Theta$ denote the resulting bijection. Fix $\theta,\theta'\in\Theta$ with $a^*(\theta)\le a^*(\theta')$. By QCC, $u(\cdot,\theta)$ and $u(\cdot,\theta')$ are unimodal with modes at $a^*(\theta)$ and $a^*(\theta')$, respectively. For each $a\in A$, if $\Delta u(a,\theta)>0$, unimodality of $u(\cdot,\theta)$ implies $a\le a^*(\theta)\le a^*(\theta')$, hence $\Delta u(a,\theta')\ge 0$. Likewise, if $\Delta u(a,\theta')<0$, then $a\ge a^*(\theta')\ge a^*(\theta)$, hence $\Delta u(a,\theta)\le 0$. Hence, along the order induced by $\beta$, the vector $\big(\Delta u(a,\theta)\big)_{\theta\in\Theta}$ is quasi‑monotone. Therefore, the decision problem is LSC.
\end{proof}

\bibliographystyle{ecta-fullname} 
\bibliography{add}  




\end{document}